\documentclass[aps,prl,twocolumn,superscriptaddress,floatfix,letter,longbibliography]{revtex4-2}

\usepackage{epsfig,amsmath,amssymb,color,amsfonts,physics,microtype}
\usepackage[bookmarks=true,colorlinks,citecolor=red]{hyperref}
\usepackage{dsfont}
\usepackage{wrapfig}
\usepackage{tikz}
\usepackage{physics}
\usepackage{mathrsfs}
\usepackage[export]{adjustbox}
\usetikzlibrary{quantikz}
\usepackage{soul}
\usepackage{comment}
\newcommand{\stab}{\mathfrak{s}}

\begin{document}

\title{Clifford-Dressed Variational Principles for Precise Loschmidt Echoes}

\author{Antonio Francesco Mello}
\affiliation{International School for Advanced Studies (SISSA), via Bonomea 265, 34136 Trieste, Italy}
\affiliation{Center for Computational Quantum Physics, Flatiron Institute, 162 5th Avenue, New York, NY 10010}
\author{Alessandro Santini}
\affiliation{International School for Advanced Studies (SISSA), via Bonomea 265, 34136 Trieste, Italy}
\affiliation{CPHT, CNRS, Ecole Polytechnique, Institut Polytechnique de Paris, 91120 Palaiseau, France}
\affiliation{Collège de France, Université PSL, 11 place Marcelin Berthelot, 75005 Paris, France}
\author{Mario Collura}
\affiliation{International School for Advanced Studies (SISSA), via Bonomea 265, 34136 Trieste, Italy}
\affiliation{INFN Sezione di Trieste, 34136 Trieste, Italy}

\begin{abstract}
We extend the recently introduced Clifford dressed Time-Dependent Variational Principle (TDVP) to efficiently compute many-body wavefunction amplitudes in the computational basis. This advancement enhances the study of Loschmidt echoes, which generally require accurate calculations of the overlap between the evolved state and the initial wavefunction.

By incorporating Clifford disentangling gates during TDVP evolution, our method effectively controls entanglement growth while keeping the computation of these amplitudes accessible.
Specifically, it reduces the problem to evaluating the overlap between a Matrix Product State (MPS) and a stabilizer state, a task that remains computationally feasible within the proposed framework.

To demonstrate the effectiveness of this approach, we first benchmark it on the one-dimensional transverse-field Ising model. We then apply it to more challenging scenarios, including a non-integrable next-to-nearest-neighbor Ising chain and a two-dimensional Ising model.

Our results highlight the versatility and efficiency of the Clifford-augmented MPS, showcasing its capability to go beyond the evaluation of simple expectation values. This makes it a powerful tool for exploring various aspects of many-body quantum dynamics.
\end{abstract}

\maketitle

\paragraph{Introduction.---} Understanding the non-equilibrium dynamics of quantum many-body systems is a central challenge in quantum physics~\cite{Alba2018SciPost,Alba2016EntanglementAT}, due to the exponential growth of the Hilbert space and the complex quantum correlations, such as entanglement, that develop during time evolution~\cite{Calabrese_2005,RevModPhys.80.517}. As entanglement increases, simulating these systems becomes progressively more difficult, demanding advanced numerical methods capable of managing this complexity~\cite{Feynman1982,Daley_2022,xu2023herculean,Pashayan2020fromestimationof}. Matrix Product States (MPS) have emerged as a powerful tool for representing one-dimensional quantum systems in a computationally efficient manner, providing a compact description of quantum states that evolves over time~\cite{Paeckel_2019, Schollwock_2011,Vidal_2004,Collura2024}.

The Time-Dependent Variational Principle (TDVP) is a widely-used method for simulating quantum dynamics by evolving the parameters of an MPS within its variational manifold~\cite{Haegeman_2016,TDVP_2011, biamonte2020lectures,Silvi_2019}. This approach allows the quantum state to remain an optimal approximation within the MPS ansatz throughout the evolution, enabling accurate long-time simulations. However, even with TDVP, the growth of entanglement during non-equilibrium processes remains a major limitation, as it leads to an ever-increasing bond dimension, which in turn raises the computational cost of maintaining an accurate representation of the system~\cite{wu2022disentanglinginteractingsystemsfermionic}.

To address this challenge, recently it was proposed a novel technique that integrates Clifford-based disentangling circuits into the TDVP framework~\cite{qian2024clifford,mello2024clifford}. Clifford circuits are classically tractable, even when the quantum states they generate exhibit substantial entanglement~\cite{Dehaene_2003,Gottesman_1998_1,Gottesman_1998_2,Gottesman_1997}. In this approach, a structured sequence of Clifford operations are used to \emph{dress} the operators such as the Hamiltonian while disentangling the quantum state.
The evaluation of expectation values of observables is quite natural within the Clifford-dressed version of TDVP. On the other hand, calculating quantities such as fidelities and amplitudes in the computational basis—which have recently been studied in the context of discrete circuit dynamics~\cite{liu2024classicalsimulabilitycliffordtcircuits}—is less straightforward.

In this work, we adapt the Clifford enhanced TDVP method designed to study Loschmidt echoes in spin systems evolving under Hamiltonian dynamics. Loschmidt echo is a fundamental concept for characterizing a wide range of phenomena in non-equilibrium dynamics~\cite{PhysRevA.94.010102,doi:10.1098/rsta.2015.0160,Heyl1,Heyl3,Heyl4,PhysRevX.11.041018,Heyl2,karch2025probingquantummanybodydynamics}, such as the work statistics of quantum systems~\cite{PhysRevLett.110.230602,PhysRevLett.110.230601} and decoherence~\cite{cucchietti2003decoherence, PhysRevA.85.060101} . The behavior of this quantity has been often investigated using numerical methods, and analytical expressions have been derived in some cases~\cite{Piroli_2017,GORIN200633,PhysRevLett.124.160603,PhysRevLett.96.140604}. Recently, it has also been measured on quantum hardware to probe the equilibrium properties of quantum systems \cite{PRXQuantum.5.030323}.

The Loschmidt amplitude is defined as the overlap between a time-evolving system and its initial state, which is typically a stabilizer state. In the Clifford-dressed version of the TDVP algorithm, the evolved state is represented as a Clifford-enhanced MPS (CMPS) with a fixed maximum bond dimension. The action of the Clifford unitaries can be back-propagated into the initial stabilizer state, transforming it into another stabilizer state.

As a result, evaluating Loschmidt amplitudes reduces to computing the overlap between an MPS and a stabilizer state. To achieve this, we employ two different techniques: the first relies on a perfect sampling strategy over the target stabilizer state, while the second approximates the projection of the MPS onto the stabilizer state using its stabilizer group generators. 

We benchmark our approaches on the one dimensional transverse field Ising model, and then investigate more complex systems, namely the next-to-nearest neighbors Ising chain and the $2$D Ising model. By carefully managing entanglement throughout the simulation, our approach paves the way for exploring quantum dynamics in regimes that were previously beyond the reach of classical methods.\\

\paragraph{Framework.---} We consider a system of $N$ qubits. The computational basis states ${\ket{0}, \ket{1}}$ are taken as the eigenstates of the Pauli matrix $\hat{\sigma}^3$, such that $\hat{\sigma}^3 \ket{s} = (-1)^s \ket{s}$. A general Pauli string on $N$ qubits can be written as $\hat{\Sigma}^{\boldsymbol{\mu}} = \hat{\sigma}^{\mu_1}_1 \hat{\sigma}^{\mu_2}_2 \cdots \hat{\sigma}^{\mu_N}_N$, where the superscript $\boldsymbol{\mu} = (\mu_1, \dots, \mu_N)$ specifies the Pauli operators $\mu_j \in {0, 1, 2, 3}$ acting on the $j$th qubit, with $\mu_j = 0$ representing the identity operator. 
Pauli strings form a complete basis for operators acting on the Hilbert space $\mathcal{H} = span \left(\lbrace{\ket{0}, \ket{1}}\rbrace^{\otimes N}\right)$. Specifically, any operator $\hat{O}$ can be expanded as $\hat{O} = \sum_{\boldsymbol{\mu}} O_{\boldsymbol{\mu}} \hat{\Sigma}^{\boldsymbol{\mu}}$, where the coefficients are given by $O_{\boldsymbol{\mu}} = 2^{-N}\mathrm{Tr}(\hat{O} \hat{\Sigma}^{\boldsymbol{\mu}})$. The Pauli strings satisfy the orthogonality condition $\mathrm{Tr}(\hat{\Sigma}^{\boldsymbol{\mu}} \hat{\Sigma}^{\boldsymbol{\nu}}) = 2^N \delta_{\boldsymbol{\mu}\boldsymbol{\nu}}$.

When an operator is conjugated by a general unitary $\hat{U}$, i.e., $\hat{U} \hat{O} \hat{U}^\dagger$, its representation in the Pauli basis can grow increasingly complex. However, if the unitary is chosen from the Clifford group $\mathcal{C}_N$, the Pauli string structure remains intact. Specifically, for any Clifford unitary $\hat{C} \in \mathcal{C}_N$ and any Pauli string $\hat{\Sigma}^{\boldsymbol{\mu}}$, we have $\hat{C} \hat{\Sigma}^{\boldsymbol{\mu}} \hat{C}^\dagger = \pm \hat{\Sigma}^{\boldsymbol{\nu}}$, where $\hat{\Sigma}^{\boldsymbol{\nu}}$ is another Pauli string. The Clifford group is generated by the Hadamard gate ($H$), phase gate ($S$), and the entangling CNOT gate (${\rm CX}$), which play a fundamental role in fault-tolerant quantum computing and quantum error correction~\cite{Bravyi_2005,Nielsen_chuang_2010,kitaev2002classical}.

Recently, Clifford transformations have been integrated into the Tensor Network formalism, leading to the development of a new class of variational wave functions known as Clifford-enhanced Matrix Product States~\cite{lami2024quantum, qian2024augmenting} or Stabilizer Tensor Networks~\cite{masotllima2024}, along with their corresponding dressed operators~\cite{mello2024hybrid}. This advancement has enabled the design of efficient algorithms for both equilibrium~\cite{qian2024augmenting,huang2024cliffordcircuitsaugmentedmatrix,qian2024augmentingfinitetemperaturetensor} and out-of-equilibrium dynamics~\cite{mello2024clifford, qian2024clifford, fux2024disentanglingunitarydynamicsclassically}. These methods are based on modifications of well-established frameworks such as the Density Matrix Renormalization Group (DMRG)~\cite{white1993density} and the Time-Dependent Variational Principle (TDVP)~\cite{Haegeman_2016}.

For the purpose of this work, let us briefly recap the Clifford-dressed TDVP algorithm. This method leverages the stabilizer formalism to construct appropriate Clifford unitaries that disentangle the MPS wave function. Specifically, at each time step $m \in \{1, 2, \dots\}$ corresponding to times $t_m = m \, dt$, an optimized transformation $C(t_m)$, composed of two-site Clifford gates, is applied to the evolved state to reduce its entanglement. The resulting state is thus given by the Clifford augmented MPS $\ket*{\tilde{\psi}(t_m)} = C(t_m) \ket{\psi(t_m)}$~\cite{ mello2024clifford, masotllima2024, qian2024augmenting, lami2024quantum, qian2024augmentingfinite}, and the Hamiltonian is transformed accordingly. Notably, being the optimized transformation $C(t_m)$ a Clifford operation, the complexity of the Hamiltonian in the Pauli basis does not increase, nor does its representation as a diagonal MPO.\\

\begin{figure}[t!]
    \centering
    \includegraphics[width=0.8\linewidth]{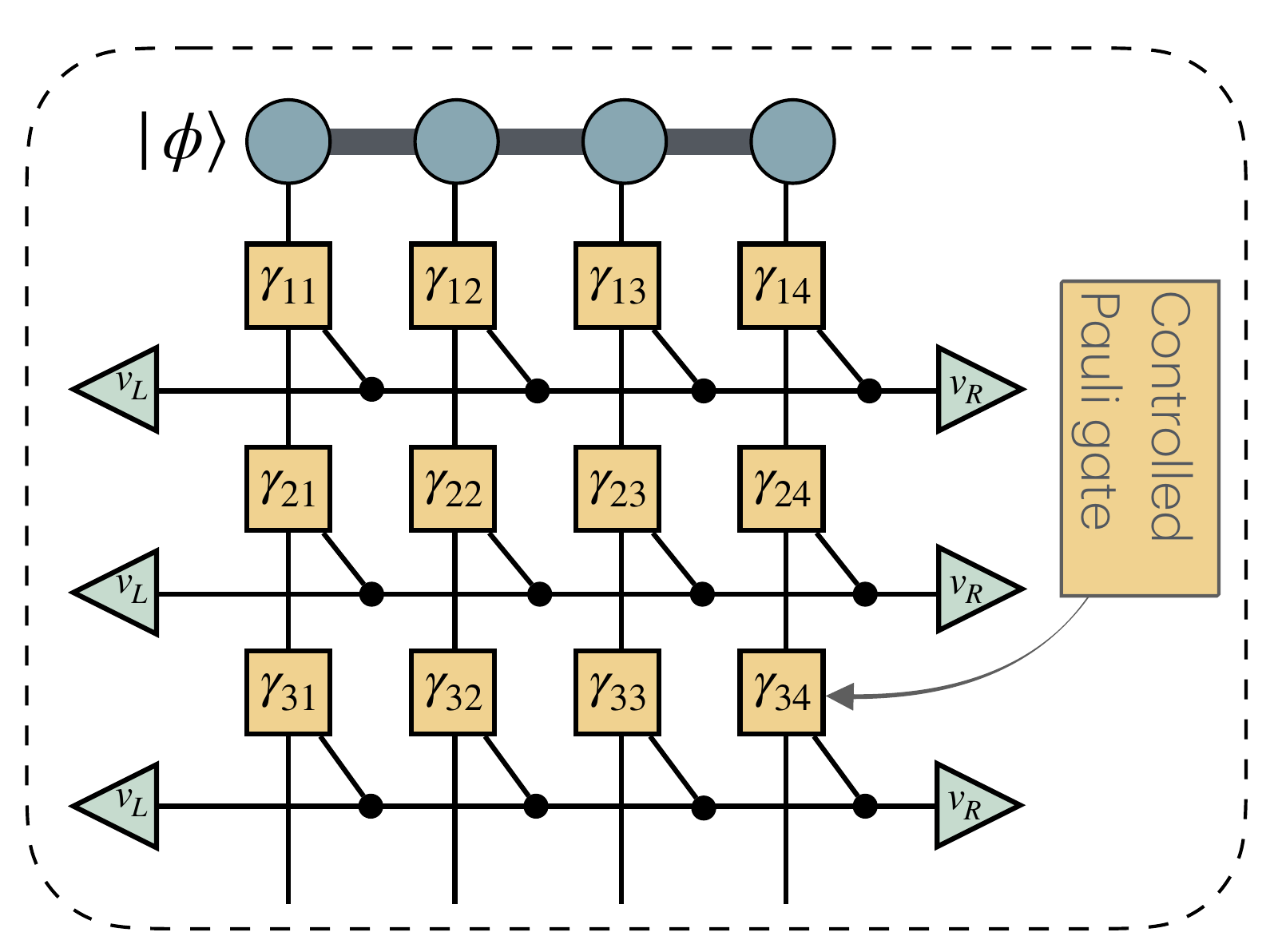}
    \caption{Tensor network representation of the projection of an MPS $\ket{\phi}$ over the stabilizer group associated to a stabilizer state $\ket{\stab}$. Each generator $P_j$ of the stabilizer group is represented as a bond dimension $2$ MPO, whose entries $G_{jk}\ {k=1,\dots, N}$ can be interpreted as controlled-Pauli gates.}
    \label{fig:projection_tn}
\end{figure}

\paragraph{Stabilizer - MPS fidelity.---} As will be explained in detail in the next section, the core task in our approach is the computation of the fidelity between a stabilizer state and an MPS. Stabilizer states, generated by Clifford circuits, are classically tractable due to their efficient representation in terms of Pauli strings. On the other hand, MPS offer a flexible variational ansatz for approximating quantum states with limited entanglement. Let us suppose that $\ket{\stab}$ is a stabilizer state whose generators are given by $\{g_1, ..., g_N\}$, such that $g_j^2 = \mathbb{I}$, and $\ket{\phi}$ is an MPS with bond dimension $\chi$. Let us also consider the bit string $\ket{\mathbf{x}}=\ket{x_1,...,x_N}$, with $x_j\in \{0,1\}$, which is trivially a stabilizer state as well, with stabilizer group generators $\{(-1)^{x_1}\sigma^{3}_1,\dots,(-1)^{x_N}\sigma^{3}_N\}$.
To compute the overlap between this two class of  states we can basically use two different approaches:\\

{\it 1.---} The first method exploits the fact that we can {\it perfectly sample} over the probability distribution of a MPS wave function~\cite{Stoudenmire_2010} or a stabilizer state wave function~\cite{PhysRevA.70.052328}. 
In particular, in order to discard all vanishing amplitude $\braket{\mathbf{x}}{\stab}$, it is convenient to sample over the stabilizer state such that the overlap can be rewritten as
\begin{equation}
    \braket{\stab}{ \phi} = \mathbb{E}_{\mathbf{x}\sim \abs{\braket{\mathbf{x}}{\stab}}^2} \left[\frac{\braket{\mathbf{x}}{\phi}}{\braket{\mathbf{x}}{\stab}}\right].
\end{equation} 
Classical configuration $\mathbf{x}$ can be extracted according to the Born rule
$\abs{\braket{\mathbf{x}}{\stab}}^2$ with a computational complexity $O(N^3)$, while
the numerator $\braket{\mathbf{x}}{\phi}$  in the square bracket requires $O(N\chi^2)$ multiplications. The variance of this stochastic estimator can be readily evaluated. Indeed, by considering $S =  \displaystyle \frac{\braket{\mathbf{x}}{\phi}}{\braket{\mathbf{x}}{\stab}}$, one can easily find that
\begin{subequations}
\begin{align}
     \overline{|S|^2}  &= \sum_{\mathbf{x}}   \left| \braket{\mathbf{x}}{\stab}\right|^2\left|\frac{\braket{\mathbf{x}}{\phi}}{\braket{\mathbf{x}}{\stab}}\right|^2 = \sum_{\mathbf{x}} |\phi(\mathbf{x})|^2 = 1 \\
    | \overline{S }| ^2 &=\left| \sum_{\mathbf{x}}  \left| \braket{\mathbf{x}} {\stab}\right|^2 \frac{\braket{\mathbf{x}}{\phi}}{\braket{\mathbf{x}}{\stab}}\right|^2 = \left|\braket{\stab}{\phi}\right|^2
\end{align}
\end{subequations}
and thus ${\rm Var}(S) = 1 - \left|\braket{\stab} {\phi}\right|^2$, where we used $\braket{\phi}{\phi}=1$. Remarkably, this result shows that the variance of the stochastic estimator is not increasing with the system size.\\

{\it 2.---} The second approach to compute $\braket{\stab}{\phi}$, which was also introduced in~\cite{liu2024classicalsimulabilitycliffordtcircuits}, is to project $\ket{\phi}$ in the stabilizer group generated by the generators $\{g_1,...,g_N\}$ where $g_j = \theta_j \sigma^{\gamma_{j1}}... \sigma^{\gamma_{jN}}$, $\theta_j =\pm 1$, and is such that $g_j\ket{\stab}=\ket{\stab}$. In fact
\begin{equation}
\ketbra{\stab} = 
\prod_{j=1}^{N} P_j
\end{equation}
where the projector $P_j$ on the $j$th generator has an MPO representation with auxiliary dimension $2$ \begin{equation}
    P_j = \frac{\mathbb{I}+g_j}{2} = \mathbf{v}_j^T \cdot {\mathbb{G}_{j1} \cdots \mathbb{G}_{jN}}\cdot \mathbf{v}_j
\end{equation}
where $\mathbb{G}_j$ is the following a Control-Pauli gate \begin{equation}
    \includegraphics[width=\linewidth,valign=c]{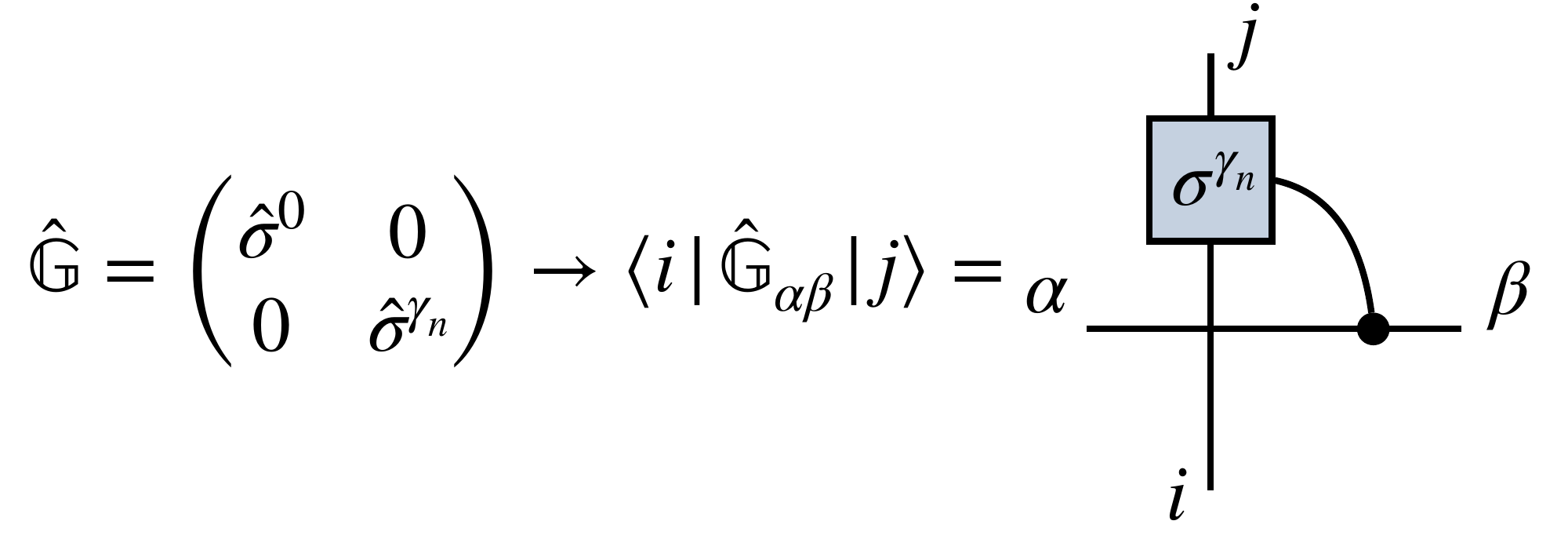},
\end{equation}
and $\mathbf{v}_j = (1/\sqrt{2},\ \sqrt{\theta_j/2})^T$ is the boundary vector.

We can therefore write the projection as the  network in Fig.~\ref{fig:projection_tn}. After applying the projector to the MPS, retrieving the amplitude over the stabilizer state $\ket{\stab}$ can be easily done by sampling a certain configuration $\ket{\mathbf{x}}$ from $\ket{\stab}$, calculating $\braket{\mathbf{x}}{\stab}\braket{\stab}{\phi}$, and then dividing by the amplitude $\braket{\mathbf{x}}{\stab}$.

Remarkably, both methods introduce a certain degree of approximation in the evaluation of the overlap. The former, being stochastic in nature, is limited by the number of samples used. The latter, instead, is affected by the application of the projector over the stabilizer group, which in principle has to be performed by repeatedly truncating the MPS bond dimension.\\

\paragraph{Loschmidt Echo with CMPS.---} We are interested in the evaluation of Loschmidt echoes for a unitarily evolved system. Specifically, we want to compute 
\begin{align}
    \mathcal{L}(t) =  |\braket{\psi(0)}{\psi(t)}|^2
\end{align}
where $\ket{\psi (0)}$ is an initial shortly correlated stabilizer state, and $\ket{\psi(t)} = U(t)\ket{\psi(0)}$, with $U(t)=e^{-iHt}$. Time evolution is well known to dramatically increase the amount of entanglement in the starting state, thus making the TN description of the system inaccurate for long times. For this reason, we employ the Clifford dressed version of the TDVP algorithm. In particular, within this framework, the evaluation of the Loschmidt amplitude at the step $m$ of evolution precisely reduces to the calculation of $\braket{\stab_m}{\phi_m}$, where $\ket{\stab_m} = C(t_m)\ket{\stab_{m-1}} = \prod_m C_m \ket{\psi(0)}$ and $\ket{\phi_m}$ is the lowly-entangled version of the evolved state. The fact that the starting state is a stabilizer state underpins the effectiveness of the whole protocol. Indeed, the stabilizer formalism provides a suitable framework to efficiently construct $\ket{\stab_m} =\prod_m C_m \ket{\psi(0)}$ provided that $\ket{\psi(0)} = \ket{\stab_0}$. While for standard MPS algorithms the reiterated application of such Clifford transformations would in general result in a dramatic increase of the bond dimension, and thus of the needed computational resources, for stabilizer states it just boils down to a sequential composition of all the tableux associated to each $C_m$. 

The requirement of a starting stabilizer state is not restrictive as it might initially seem. As a matter of fact, the proposed protocol applies to all the states of the form $C\ket{0}^{\otimes N}$, with $C \in \mathcal{C}_N$, that admit an MPS representation. 

\begin{figure}[t!]
    \centering
    \includegraphics[width=\linewidth]{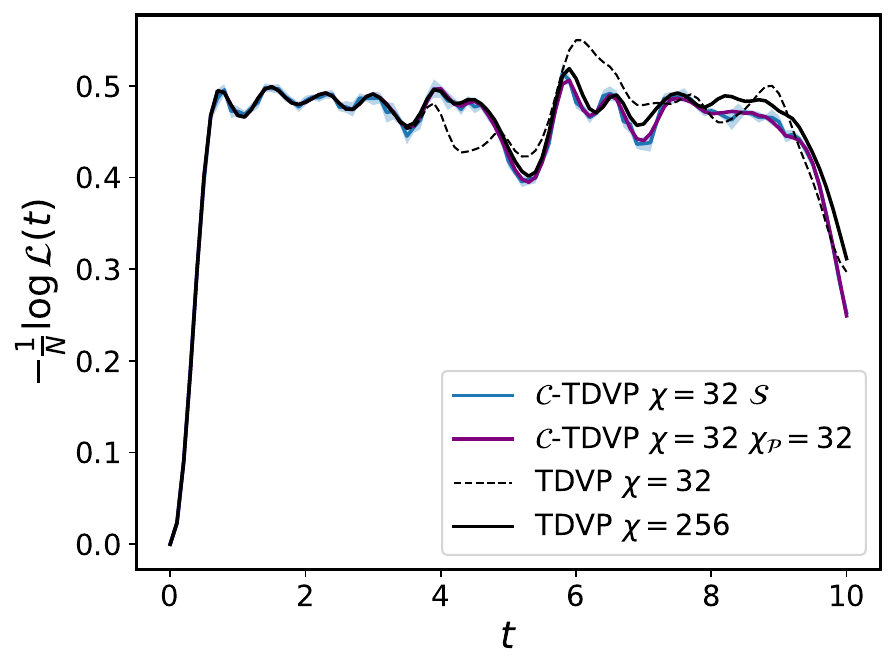}
    \caption{Rescaled Loschmidt echo for $N=20$ spins prepared in the state $\ket{0}^{\otimes N}$ and evolved through the critical Ising Hamiltonian. Both the $\mathcal{C}-$TDVP techniques (stochastic blue line, projective purple line) outperform the standard TDVP algorithm (dashed grey line) at fixed bond dimension $\chi=32$, and approach the performance it has for higher values of $\chi=256$ (black line). For the stochastic method, averages are taken over $10^4$ samples. In the legend, $\chi_\mathcal{P}$ refers to the maximum bond dimension during the stabilizer group projection. The time step size has been set to $dt=0.1$.}
    \label{fig:loschmidt}
\end{figure}

We benchmark the introduced protocols on an $N=20$ qubits system prepared in the stabilizer state $\ket{\psi(0)}=\ket{0}^{\otimes N}$, and evolved through the Ising Hamiltonian 
\begin{align}
    H = \sum^{N-1}_{i=1}\sigma^1_i\sigma^1_{i+1} -h \sum^N_{i=1}\sigma^3_i
\end{align}\label{eq:ising_hamiltonian}
at the critical point $h=h_c=1$. As it can be seen from Fig.~\ref{fig:loschmidt}, where we plot the rescaled Loschmidt echo for better visualization, both the proposed methods within the Clifford-dressed TDVP outperform the standard TDVP routine, yielding more accurate values of the Loschmidt amplitudes. Indeed, at fixed amount of resources $\chi = 32$, $\mathcal{C-}$TDVP reaches an advantage over TDVP of about two time units.

We now study the extent of applicability of such techniques in the evaluation of Loschmidt amplitudes for more challenging systems. In the following, we resort to method $2$ for the evaluation of the overlaps.\\

\begin{figure}[t!]
    \centering
    \includegraphics[width=\linewidth]{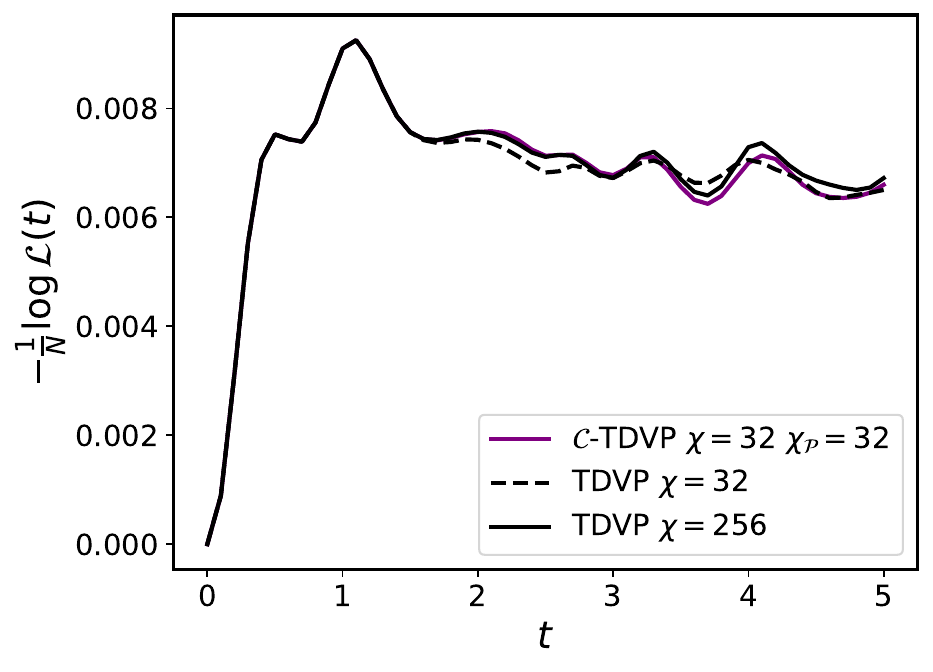}
    \caption{Rescaled Loschmidt echo for $N=20$ spins prepared in the state $\ket{0}^{\otimes N}$ and evolved through the non-integrable next-to-nearest neighbors Ising Hamiltonian with transverse field $h=1$. 
     In the legend, $\chi$ refers to the fixed bond dimension used during the $1-$site tdvp update, while $\chi_{\mathcal{P}}$ represents the maximum bond dimension during the stabilizer group projection. The time step size has been set to $dt=0.1$.}
    \label{fig:loschmidt_NNN}
\end{figure}

\paragraph{Clifford-Enhanced Loschmidt Echo in NNN and 2$D$ Ising Models.---} The first scenario we consider is that of an $N=20$ qubits system prepared in the state $\ket{\psi(0)}=\ket{0}^{\otimes N}$, and evolved through the non-integrable Ising Hamiltonian with next to nearest neighbors interactions 
\begin{align}
H = \sum^{N-1}_{i=1}\sigma^1_i\sigma^1_{i+1} + \sum^{N-2}_{i=1}\sigma^1_i\sigma^1_{i+2} - h\sum^N_{i=1}\sigma^3_i    
\end{align}
 and transverse field $h=1$. From Fig.~\ref{fig:loschmidt_NNN} it can be noticed that also in this case the $\mathcal{C}-$TDVP evaluation of Loschmidt amplitudes leads to a two time units gain with respect to standard TDVP with the same bond dimension $\chi$. 

\begin{figure}[t!]
    \centering
    \includegraphics[width=\linewidth]{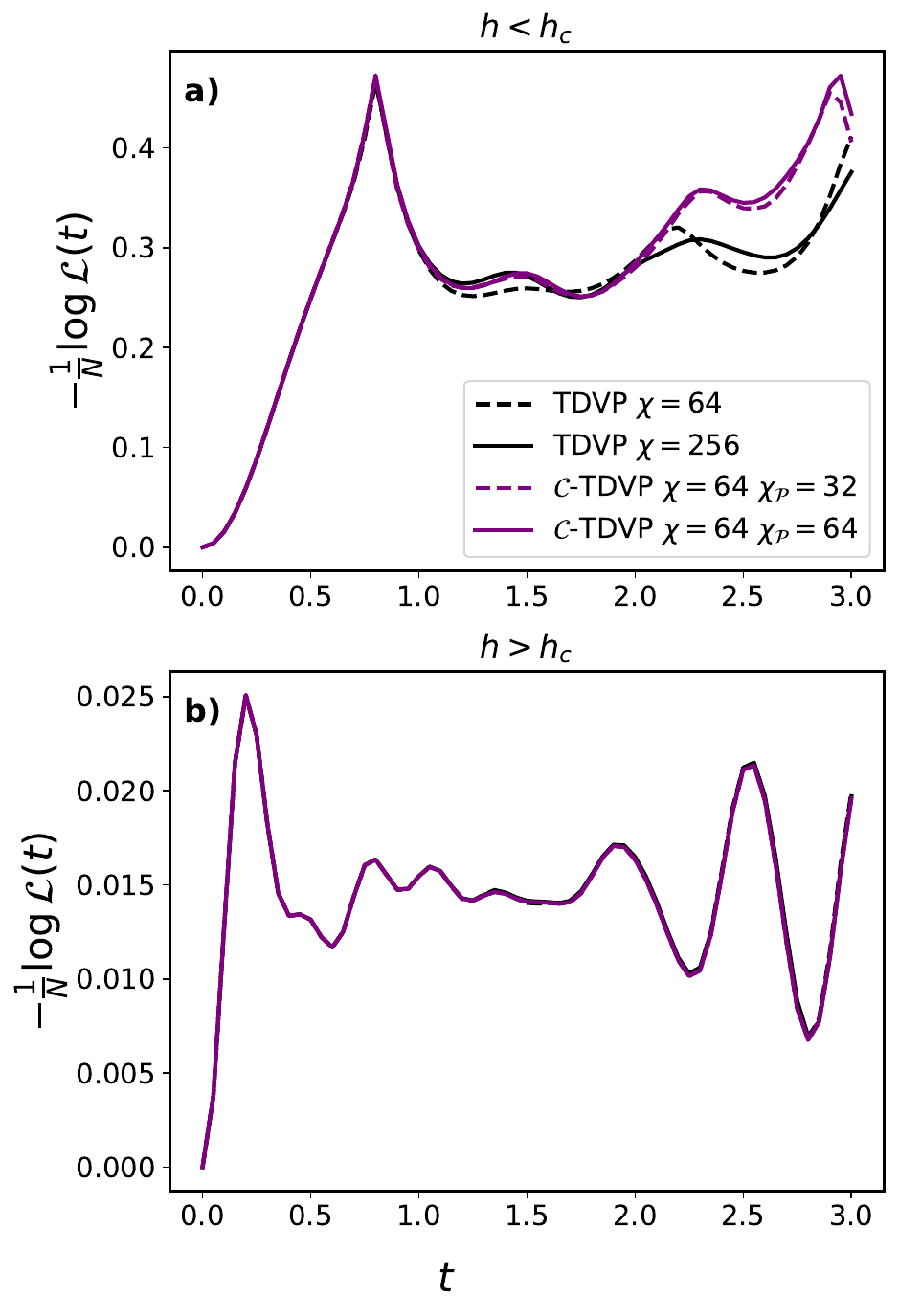}
    \caption{Rescaled Loschmidt echo for a 2D system of size $L=5$. In panels a) and b) the state $\ket{0}^{\otimes N}$ is evolved through the Ising Hamiltonian with transverse field $h=1 <h_c$ and $h=4>h_c$ respectively.
     In the legend, $\chi$ refers to the fixed bond dimension used during the $1-$site tdvp update, while $\chi_{\mathcal{P}}$ represents the maximum bond dimension during the stabilizer group projection. The time step size has been set to $dt=0.05$.}
    \label{fig:loschmidt2D_ferro}
\end{figure}

We now push the analysis further by studying a $2$D system evolved through the nearest-neighbor Ising Hamiltonian
\begin{equation}
    H = \sum_{\langle i,j \rangle} \sigma_i^1 \sigma_j^1 - h \sum_i \sigma_i^3
\end{equation}
on a square lattice with open boundary conditions. This model displays a quantum phase transition for transverse field $h=h_c=3.044$. Specifically, we consider a $5 \times 5$ system prepared in the state $\ket{\psi(0)}=\ket{0}^{\otimes{N}}$. We employ an MPS representation of the two-dimensional state, constructed by unfolding the square lattice along its diagonals. Fig.~\ref{fig:loschmidt2D_ferro} displays the results both for $h<h_c$ and $h>h_c$ in panel a) and b) respectively.
The case $h>h_c$ turns out to be the less interesting one, as both the Clifford-dressed and the plain TDVP perform quite well with the same amount of resources. For $h<h_c$, instead, the $\mathcal{C}-$TDVP happens to give slightly more accurate result for about one time unit. Remarkably, we see that the maximum bond dimension employed during the stabilizer group projection can be reduced with respect to the one used while time evolving the state without necessarily loosing accuracy in the overlap evaluation. 


In the case of the $2$D Ising model, it is also interesting to further analyze the efficiency of the Clifford disentangling procedure during time evolution. Indeed,~\cite{mello2024clifford} is entirely focused on $1$D systems, while in~\cite{qian2024clifford} the authors use the $2-$sites TDVP scheme on different models. 
\begin{figure}[t!]
    \centering
    \includegraphics[width=\linewidth]{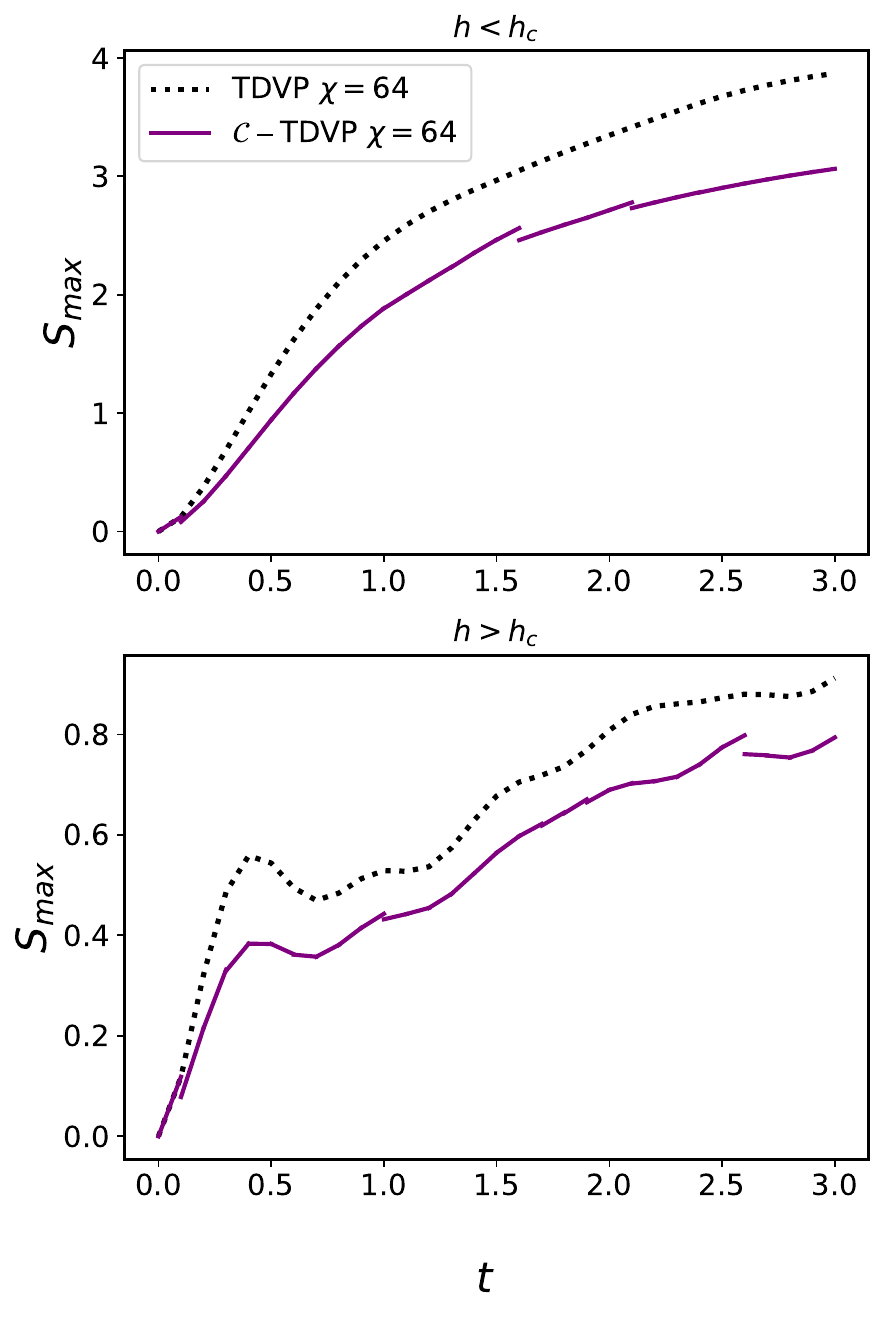}
    \caption{Maximum entanglement entropy for a $5 \times 5$ square lattice prepared in the state $\ket{0}^{\otimes N}$ and evolved through the Ising Hamiltonian. Left panel: $h=1<h_c$. Right panel: $h=4>h_c$.}
    \label{fig:ent_entropy2D}
\end{figure}
Fig.~\ref{fig:ent_entropy2D} shows the entanglement entropy growth during time evolution. The upper and lower panels refer to the $h<h_c$ and $h>h_c$ cases respectively. In both scenarios, the Clifford disentangling routine is effectively managing to reduce the entanglement content of the state, although not at every time step. In the $h > h_c$ setup, this does not lead to improved performance at a fixed $\chi$, as the entanglement remains low on the analyzed time scale. In contrast, for the $h < h_c$ case, the disentangling routine improves the performance at a fixed amount of resources when evaluating the Loschmidt amplitudes.\\

\paragraph{Conclusions.---} In this work, we leveraged the Clifford-dressed version of the TDVP algorithm for MPS to evaluate the Loschmidt amplitudes for spin systems. Specifically, we presented two methods to efficiently evaluate overlaps between stabilizer states and MPS and applied them to study the non-equilibrium dynamics of many-body quantum systems.

We benchmarked our protocols on the $1$D transverse-field Ising model, and then analyzed the extent of applicability to the non-integrable next-to-nearest neighbors Ising chain along with the $2$D Ising model on the square lattice. Across all scenarios, we found that incorporating Clifford-disentangling techniques within standard TDVP enhances the time reach of numerical simulations for evaluating Loschmidt echoes at a fixed amount of computational resources.

Our work extends the possible applications of the Clifford-dressed TDVP algorithm beyond the evaluation of observables, showing its broader relevance in the study of out-of-equilibrium dynamics.\\

Note Added: While finalizing this manuscript, the work~\cite{liu2024classicalsimulabilitycliffordtcircuits} was published on the arXiv. In this work, the authors independently employ the same algorithm that we presented as method {\it 2} to sample a CMPS in the computational basis.\\

\paragraph{Acknowledgments.---} We acknowledge the use of Stim \cite{stim} for the stabilzer formalism operations and ITensor for the TN simulations \cite{itensor}. We are particularly grateful to G. Lami, J. De Nardis, M. Dalmonte, E. Tirrito for collaborations on topics connected with this work. This work was supported by the PNRR MUR project PE0000023-NQSTI, and by the PRIN 2022 (2022R35ZBF) - PE2 - ``ManyQLowD''. The Flatiron Institute is a division of the Simons Foundation.

\bibliography{bib}

\end{document}